\documentclass[12pt, a4paper]{article}
\usepackage{amsmath, amsfonts}%, ntheorem}
\usepackage{cite}
\begin{document}

% \theoremstyle{nonumberplain}
% \theorembodyfont{\rmfamily}
% \theoremseparator{:}
% \theoremsymbol{$\blacksquare$}
% \newtheorem{proof}{Proof}

% \theoremstyle{plain}
\newtheorem{definition}{Definition}
\newtheorem{theorem}{Theorem}
\newtheorem{lemma}{Lemma}

\title{On Lower Bound of Worst Case Error Probability for Quantum Fingerprinting with Shared Entanglement}

\author{Xin Li, Tian Liu, Han Peng\footnote{Corresponding author, penghan@tcs.pku.edu.cn}, Hongtao Sun, Jiaqi Zhu\\ \it School of EECS, Peking University, P.R.China}

  \maketitle
  \begin{abstract}
  This paper discusses properties of quantum fingerprinting with shared entanglement. Under certain restriction of final measurement, a relation is given between unitary operations of two parties. Then, by reducing to spherical coding problem, this paper gives a lower bound of worst case error probability for quantum fingerprinting with shared entanglement, showing a relation between worst case error probability and the amount of entanglement(measured by Schmidt number).
  \end{abstract}

\maketitle
\section{introduction}
Fingerprinting is a useful tool to determine whether two copies of message are identical, especially for large and distributed data where both parties have to communicate to accomplish this task. Instead of comparing original messages, it is enough to compare short messages called "fingerprints" of them if some manageable error probability is allowed. The goal of fingerprinting is to reduce the amount of communication while maintain acceptable error probability.

The notion of fingerprinting first appeared in communication complexity (\cite{KUSHILEVITZ}) research where the amount of communication is calculated and local computational complexity ignored. Discussion in this paper is based on SMP (Simultaneous Message Passing) model introduced by Yao in \cite{YAO79}, where 3 distributed parties, say Alice, Bob, and Roger, are involved. They cooperate to compute some function $f(x, y)$, but Alice holds $x$, Bob holds $y$. Mutual communication is not allowed between Alice and Bob, but they can each send one message to Roger, who will then compute $f(x, y)$ based on received messages. This problem can be trivially solved if Alice and Bob send entire messages to Roger, but the question is: what error probabilty can be achieved if they transfer shorter messages? When $f(x, y)$ is defined as
\begin{equation}\label{intro:def_f}
f(x, y) = \left\{\begin{array}{ll}
		1 & \textrm{if $x = y$}\\
		0 & \textrm{else}
		\end{array}
		\right.
\end{equation}
the messages sent by Alice and Bob are called \emph{fingerprints} of $x$ and $y$, respectively. When qubits are used as fingerprints, they are called quantum fingerprints.

Quantum fingerprinting was introduced in \cite{BCWdW}, where the author showed that $O(\log{n})$ qubits of communication suffice to determine the equality of two $n$-bit strings with bounded error probability. On the other hand, however, Ambainis proved in \cite{AMBAINIS96} that $\Theta(\sqrt{n})$ classical bits are necessary for the same problem, showing an exponential gap. Later in \cite{SCOTT} Scott et. al. analyzed the relation between worst case error probability and actual number of transferred qubits, giving a lower bound on achievable worst case error probability.

However, the first quantum fingerprinting scheme in \cite{BCWdW} does not consider shared \emph{a prioi} entanglement between Alice and Bob. In \cite{HORN}, Horn et. al. discussed such case. Since shared entanglement can easily be used as shared randomness, the authors also made comparison between classical fingerprinting with shared randomness and quantum fingerprinting with shared entanglement.

This paper can be seen as an extension of \cite{SCOTT} onto the shared entanglement case of quantum fingerprinting. We discuss properties of shared entanglement quantum fingerprinting under certain restriction of measurement, and give a relation between lower bound of achievable worst case error probability and amount of entanglement(measured by Schmidt number).

\section{Communication Model}

This section will describe the communication model of shared-entanglement quantum fingerprinting used in this paper. First introduce the definition of one-sided error.
\begin{definition}
A fingerprinting scheme is one-sided error if for any $x = y \in M$, Roger always produces correct answer.
\end{definition}

\begin{definition}
A quantum fingerprinting scheme with shared entanglement operates as follows. First, Alice and Bob receive messages $x, y \in M$, respectively. Then each of them performs certain unitary operation on their own part of shared entanglement, and sends these qubits to Roger. On receiving qubits from both parties, Roger performs a measurement and produces an output based on result. Such a scheme is characterized by 4 components:
\begin{itemize}
\item Shared entanglement $|E\rangle$, where we suppose the Schmidt decomposition of $|E\rangle$ is
\[
|E\rangle = \sum_{k=1}^{N_s}{\lambda_k |k\rangle_A|k\rangle_B}
\]
$\lambda_k$ being the Schmidt coefficients and $N_s$ the Schmidt number.

\item Alice's unitary operation set $\{U_x\}_{x \in M}$, where she applies $U_x$ on her shared entanglement upon receiving message $x$.

\item Bob's unitary operation set $\{V_x\}_{x \in M}$, where he applies $V_x$ on his shared entanglement upon receiving message $x$.

\item Roger's measurement and decision strategy, which is restricted due to the following reasons:
\begin{itemize}
\item The scheme has only one-sided error, so that Roger's decision strategy must be deterministic.

\item Roger performs such kind of POVM on received qubits:
\[
\{|\alpha\rangle\langle\alpha|, I - |\alpha\rangle\langle\alpha|\}
\]
and annouces $x = y$ if the measurement result corresponds to $|\alpha\rangle\langle\alpha|$.
\end{itemize}

\end{itemize}

\end{definition}

Define worst case error probability as
\begin{definition}
\[
P_{wce} = \max_{x,y \in M}{P\{x \ne y,\textrm{~but Roger announces } x = y\}}
\]
\end{definition}

\section{Relation of unitary operations}
As described in previous section, the Schmidt decomposition of shared entanglement between Alice and Bob is
\[
|E\rangle = \sum_{i=1}^{N_s}{\lambda_i|i\rangle_A|i\rangle_B}
\]
where $|i\rangle_A$, $i=1,2,\cdots,N_s$, are a group of orthonormal vectors in Alice's subsystem, and $|i\rangle_B, i=1,2,\cdots,N_s$, are a group of orthonormal vectors in Bob's subsystem. Extend both groups to be orthonormal bases of the two subsystems, respectively, and denote them by $|i\rangle_A, i=1,2,\cdots,n$, and $|j\rangle_B, j=1,2,\cdots,n$.

In Roger's POVM, let
\[
|\alpha\rangle = \sum_{i,j=1}^{n}{\alpha_{i,j}|i\rangle_A|j\rangle_B}
\]
where $\sum_{i,j=1}^{n}{|\alpha_{i,j}|^2} = 1$

The following lemma is easily found in linear algebra textbooks.
\begin{lemma}[\cite{NIELSEN}]\label{lemma:trace_ip}
Denote by $L_{V}^{n}$ the set of linear operators on $\mathbb{C}^n$, and define function $(\cdot, \cdot):L_{V}^{n}\times L_{V}^{n} \to \mathbb{C}$ as
\[
(A, B) \equiv tr(A^\dagger B)
\]
then $(\cdot, \cdot)$ is an inner product on $L_{V}^{n}$, which is then an $n^2$-dimension Hilbert space.
\end{lemma}

The following theorem shows the relation between unitary operations of Alice and Bob.
\begin{theorem}
$\forall x \in M, U_xKV_x^T = S$, where $S$ is constant with respect to a global constant $e^{i\theta}$ and is independent with $x$. $K$ depends only on the Schmidt decomposition of $|E\rangle$.
\end{theorem}

% \begin{proof}
\noindent{\textbf{Proof}}
Suppose Alice and Bob receive messages $x$ and $y$, respectively, then Roger receives from them the following state
\[
U_x\otimes V_y\sum_{k=1}^{N_s}{\lambda_k|k\rangle_A|k\rangle_B}
\]
Since Roger uses POVM $\{|\alpha\rangle\langle\alpha|, I - |\alpha\rangle\langle\alpha| \}$, measurement result corresponding to $|\alpha\rangle\langle\alpha|$ appears with probability
\[
q = \left|\langle\alpha|U_x\otimes V_y \sum_{k=1}^{N_s}{\lambda_k|k\rangle_A|k\rangle_B} \right|^2
\]
then we have
\[
\begin{array}{rcl}
q & = & \left|  \sum_{i,j = 1}^{n}{\alpha_{ij}^*\langle i|_A\langle j|_B} U_x \otimes V_y \sum_{k = 1}^{N_s}{\lambda_k |k\rangle_A |k\rangle_B} \right|^2\\
  & = & \left| \sum_{i,j = 1}^{n} \sum_{k=1}^{N_s} \alpha_{ij}^* \lambda_k \langle i|_A U_x |k\rangle_A \langle k|_B V_y^T |j\rangle_B    \right|^2\\
  & = & \left| \sum_{i,j = 1}^{n} \alpha_{ij}^* \langle i|_A K V_y^T|j\rangle_B \right|^2\\
  & = & \left| tr(A^\dagger U_x K V_y^T) \right|^2\\
\end{array}
\]
where $K = \sum_{k=1}^{N_S}{|k\rangle\langle k|}$. Let$S(x,y) = U_xKV_y^T$, then
\[
q = \left| tr\big(A^\dagger S(x,y)\big)         \right|^2
\]
Recall that we require this scheme has single-sided error probability, which means that when $x = y$ Roger must have $|q| = 1$, that is
\begin{equation}\label{relation:trace1}
\left| tr\big(A^\dagger S(x,x) \big) \right| = 1
\end{equation}
According to Lemma \ref{lemma:trace_ip}, Eq (\ref{relation:trace1}) is equivalent to the condition that the projection of vector $S(x,y) \in L_V^n$ onto vector $A^\dagger \in L_V^n$ has length 1. On the other hand, $A^\dagger$ itself has length 1, and $S(x,y)$ also has length 1:
\[
\begin{array}{rcl}
|S(x,y)|^2 & = & tr\Big(\big(S(x,y)\big)^\dagger S(x,y) \Big)\\
           & = & tr\big((U_xKV_y^T)^\dagger(U_xKV_y^T)\big)\\
	   & = & tr(V_y^*K^\dagger KV_y^T)\\
	   & = & tr(V_yK^\dagger KV_y^\dagger)\\
	   & = & 1\\
\end{array}
\]
Therefore, in order to satisfy (\ref{relation:trace1}), it is necessary that
\begin{equation}\label{relation:A_and_S}
A = e^{i\theta}S(x,x)
\end{equation}
where $\theta$ is an arbitrary constant, $i$ the imaginary unit. Because entries in $A$ are completely determined by $|\alpha\rangle$ in Roger's POVM, which should be constant, we have
\begin{equation}\label{relation:UVconst}
e^{i\theta}U_xKV_x^T = const
\end{equation}
\begin{flushright}
$\square$
\end{flushright}
% \end{proof}

Note that when $N_s = n\textrm{(the dimension of each subsystem)}$, $K$ has full rank and $K^{-1}$ exists, then $V_x$ is determined by $U_x$ with respect to a unit global coefficient, and vice versa
\[
V_x = (e^{-i\theta}K^{-1}U_{x}^{\dagger}A)^T
\]

\section{Lower bound of worst case error probability}
\begin{theorem}\label{lb}
$P_{wce} \ge \frac{m-N_s^2}{N_s^2(m-1)}$
\end{theorem}

\noindent{\textbf{Proof}}
$q$ is error probability when $x \ne y$
\[
\begin{array}{rcl}
q & = & \left|tr\big(A^\dagger S(x,y)\big) \right|^2\\
  & = & \left|tr\big(S(x,x) S(x,y)\big) \right|^2\\
  & = & \left|tr\big((U_xKV_x^T)^\dagger U_xKV_y^T\big)\right|^2\\
  & = & \left|tr(V_x^*K^\dagger KV_y^T) \right|^2\\
  & = & \left|tr(V_y K K^\dagger V_x^\dagger) \right|^2\\
\end{array}
\]
then
\begin{equation}\label{Pwce}
P_{wce} = \min_{|V|=m} \max_{V_x, V_y \in V, V_x \ne V_y} \left|tr(V_y K K^\dagger V_x^\dagger)\right|^2
\end{equation}
Let $W_x = V_xK, W_y = V_yK$, the above equation can be reformulated as
\[
P_{wce} = \min_{|V|=m} \max_{V_x, V_y \in V} \left|tr(W_y W_x^\dagger)\right|^2
\]
If $N_s = n$, $K$ has full rank, $W_x$~and~$W_y$~ are linear operators in $L_V^n$. According to Lemma \ref{lemma:trace_ip}, the above min-max problem can be seen as a sperical coding problem, by Welch Bound(\cite{WELCH74})
\[
P_{wce} \ge \frac{m - N_s^2}{N_s^2(m-1)}
\]
this proves the $N_s = n$ case of Theorem \ref{lb}. Now we focus on the case of $N_s < n$.

Let $J = \sum_{i=1}^{N_s}{\sqrt{\lambda_i}|i\rangle_A \langle i|_B}$, then (\ref{relation:UVconst}) can be reformulated as (where we omit global coefficient $e^{i\theta}$)
\[
U_x J (V_xJ)^T = const
\]
Choose an arbitrary basis and write $U_x$ as a matrix. Because $U_x$ is unitary, $n$ columns of this matrix represent a group of orthonormal vectors on $\mathbb{C}^n$. Denote them as $\{|u_{x,i}\rangle\}_{i = 1,2,\cdots,n}$, then
\begin{equation}\label{eq:UxJ}
U_xJ = \left(\sqrt{\lambda_1}|u_{x,1}\rangle, \sqrt{\lambda_2}|u_{x,2}\rangle, \cdots, \sqrt{\lambda_{N_s}}|u_{x,N_s}\rangle, 0, \cdots, 0\right)
\end{equation}
and similarly
\begin{equation}\label{eq:VxJ}
V_xJ = \left(\sqrt{\lambda_1}|v_{x,1}\rangle, \sqrt{\lambda_2}|v_{x,2}\rangle, \cdots, \sqrt{\lambda_{N_s}}|v_{x,N_s}\rangle, 0, \cdots, 0\right)
\end{equation}
then (\ref{relation:UVconst}) becomes
\begin{equation}\label{eq:UxJVxJT}
const = U_xJ(V_xJ)^T = \sum_{i=1}^{N_s}{\lambda_i|u_{x,i}\rangle\langle v_{x,i}|^*}
\end{equation}
similarly for $y \ne x, y \in M$
\begin{equation}\label{eq:UyJVyJT}
const = U_yJ(V_yJ)^T = \sum_{i=1}^{N_s}{\lambda_i|u_{y,i}\rangle\langle v_{y,i}|^*}
\end{equation}
Eq(\ref{eq:UxJVxJT}) and Eq(\ref{eq:UyJVyJT}) has equal constant on their left side, therefore
\[
\sum_{i=1}^{N_s}{\lambda_i|u_{x,i}\rangle\langle v_{x,i}|^*} =  \sum_{i=1}^{N_s}{\lambda_i|u_{y,i}\rangle\langle v_{y,i}|^*}
\]
then
\[
\begin{array}{rcl}
\lambda_j |u_{x,j}\rangle & = & \sum_{i=1}^{N_s}{\lambda_i |u_{y,i}\rangle\langle v_{y,i}|^*|v_{x,j}\rangle^*}\\
 & = & \sum_{i=1}^{N_s}{\lambda_i \langle v_{x,j}|v_{y,i}\rangle |u_{y,i}\rangle}
\end{array}
\]
so
\begin{equation}\label{chap5:eq:UxVxrelation}
|u_{x,j}\rangle = \sum_{i=1}^{N_s}{\frac{\lambda_i}{\lambda_j} \langle v_{x,j}|v_{y,i} \rangle |u_{y,i} \rangle}
\end{equation}
which shows that the left $N_s$ columns of $U_x$ are linear combinations of left $N_s$ columns in another $U_y$. Because $x, y$ are arbitrarily chosen messages from $M$, the left $N_s$ columns of all $U_x \in \{U_i\}_{i \in M}$ span the same linear space $\mathbb{C}^{N_s}$. Similar statement also holds for $\{V_i\}_{i \in M}$.

Now rewrite Eq(\ref{Pwce}) according to Eq(\ref{eq:VxJ})
\[
\begin{array}{rcl}
P_{wce} & = & \min_{|V| = m} \max_{V_x,V_y \in V, V_x \neq V_y} \left|\sum_{i=1}^{N_s}{\lambda_i^2 \langle v_{x,i}|v_{y,i}\rangle}   \right|^2\\
 & = & \min_{|V| = m} \max_{V_x,V_y \in V, V_x \neq V_y} \left|tr\Big(\sum_{i=1}^{N_s}{\lambda_i |v_{x,i}\rangle \langle i|} \big(\sum_{i=1}^{N_s}{\lambda_i |v_{y,i}\rangle \langle i|}\big)^\dagger\Big) \right|^2\\
 & = & \min_{|V| = m} \max_{V_x,V_y \in V, V_x \neq V_y} \left|tr\Big(\big(\sum_{i=1}^{N_s}{\lambda_i |v_{y,i}\rangle \langle i|}\big)^\dagger \sum_{i=1}^{N_s}{\lambda_i |v_{x,i}\rangle \langle i|}\Big) \right|^2\\
\end{array}
\]
Let $Q_x = \sum_{i=1}^{N_s}{\lambda_i |v_{x,i}\rangle \langle i|}$, $Q_y = \sum_{i=1}^{N_s}{\lambda_i |v_{y,i}\rangle \langle i|}$, the above equation is equivalent to
\[
P_{wce} = \min_{|V| = m} \max_{V_x,V_y \in V, V_x \neq V_y} \left| tr (Q_y^\dagger Q_x) \right|^2
\]
Note that $Q_x, Q_y$ are both linear operators on $\mathbb{C}^{N_s}$, according to Lemma (\ref{lemma:trace_ip}) and Welch Bound we have
\[
P_{wce} \ge \frac{m-N_s^2}{N_s^2(m-1)}
\]
which proves the $N_s < n$ case of Theorem \ref{lb}.

\begin{flushright}
$\square$
\end{flushright}

\section{Conclusion}
This paper discusses the relation between Alice's and Bob's unitary operators in the model of quantum fingerprinting with shared entanglement, and gives a lower bound on the worst case error probability which profiles the ability of this fingerprinting model. In this lower bound, howerver, quantity of entanglement is measured by Schmidt number. Future work includes the lower bound based on Schmidt coefficients of shared entanglement.

Moreover, Roger's POVM is restrict to be the form $\{|\alpha\rangle\langle\alpha|, I - |\alpha\rangle\langle\alpha|\}$, lower bound with more general POVM is also an open question.

%\bibliographystyle{unsrt}
%\bibliography{reference}

%\end{document}

\end{document}